\documentclass[seceq]{ptptex}
\usepackage{overcite}
\usepackage{bm}
\usepackage{graphicx}
\usepackage{wrapft}
\usepackage{here}
\usepackage{longtable}
\usepackage{supertabular}

\title{Effective-Range Expansion of the Neutron-Deuteron Scattering
Studied by a Quark-Model Nonlocal Gaussian Potential}

\author{Kenji \textsc{Fukukawa} and Yoshikazu \textsc{Fujiwara}\\}
\inst{Department of Physics, Kyoto University, Kyoto 606-8502, Japan}

\abst{The $S$-wave effective range parameters of the neutron-deuteron
($nd$) scattering are derived in the Faddeev formalism, using a nonlocal
Gaussian potential based on the quark-model baryon-baryon interaction
fss2. The spin-doublet low-energy eigenphase shift is sufficiently
attractive to reproduce predictions by the AV18 plus Urbana three-nucleon
force, yielding the observed value of the scattering length $^{2}a_{nd}$
and the correct differential cross sections below the deuteron breakup
threshold. This conclusion is consistent with the previous result for
the triton binding energy, which is nearly reproduced by fss2 without
reinforcing it with the three-nucleon force.}

\PTPindex{205}

\begin{document}
\maketitle

\section{Introduction}
Few-nucleon systems are best suited to study the underlying nucleon-nucleon
($NN$) interaction and its extension to few-nucleon forces, since many
sophisticated techniques to solve the systems yield equivalent results
that can be compared with ample experimental data.\cite{Hu95,bench01}
In fact, the three-nucleon ($3N$) system is already solved with many
realistic meson-exchange potentials, yielding insufficient binding
energies of the triton missing 0.5 to 1 MeV without the $3N$ force.\cite{No00}
An attempt to reproduce the $NN$ and $3N$ data consistently is pursued by
using the chiral effective field theory, but a complete reproduction of all
the $3N$ data in this approach is still beyond away.\cite{Ep01,Ep02} We
have applied the quark-model (QM) baryon-baryon (BB) interaction
\cite{PPNP,fss2} to the triton and hypertriton in the Faddeev formalism
and obtained many interesting results \cite{triton1,PANIC02,triton2,hypt}.
The most recent model fss2 gives a nearly correct binding energy of the
triton with the correct root-mean-square radius, preserving the sufficient
strength of the tensor force for the deuteron and the correct $^{1}S_{0}$
$NN$ scattering length.\cite{triton1} This result indicates that fss2 is
sufficiently attractive in the $^{2}S_{1/2}$ channel of the $3N$ system
without the $3N$ force. In this channel, the deuteron distortion effect
related to the strong short-range repulsion of the $NN$ interaction is
very important. In the QM BB interaction, this part is mainly described
by the quark-exchange nonlocal kernel of the one-gluon exchange
Fermi-Breit interaction, which has quite different off-shell properties
from the phenomenological repulsive core described by local potentials
in the standard meson-exchange models. The nonlocal effect resulting
from the exact antisymmetrization of six quarks is very important
to reproduce the nearly correct triton binding energy. 

The QM BB interaction is constructed for two three-quark clusters in the
framework of the resonating-group method (RGM). It is characterized not
only by the dominant nonlocality from the interaction kernel, but also
by the energy dependence originating from the normalization kernel. In the
original evaluation of the triton and hypertriton binding energies, this
energy dependence is determined self-consistently by calculating the
expectation value of the two-cluster Hamiltonian with the square integrable
three-cluster wave function.\cite{triton1,PANIC02,hypt}.
This prescription is however not applicable to the scattering problem,
since the scattering wave function is not square integrable. In the final
form of the triton and hypertriton Faddeev calculations,\cite{triton2}
the energy dependence of the RGM kernel is eliminated by a standard
off-shell transformation, using the square root of the normalization
kernel.\cite{Su08} An extra nonlocality emerges from this off-shell
transformation as a result of eliminating the energy dependence of
the RGM kernel. It is shown in Ref.\,\citen{triton2} that this
renormalized RGM prescription gives a slightly less attractive effect
to the triton and hypertriton binding energies, in comparison with
the previous self-consistent treatment. The 350 keV deficiency of the
triton binding energy predicted by fss2 after the charge dependence
correction of the $NN$ interaction is still much smaller than 0.5 to 1
MeV given by the standard meson-exchange potentials. It is therefore
interesting to examine the nonlocal effect of the QM $NN$ interaction
to the $3N$ scattering observables, such as the scattering lengths,
the differential cross sections and the spin polarization, in this
energy-independent QM $NN$ interaction.

The numerical calculation of the three-body scattering using the QM BB
interaction is very time consuming due to the complex structure of the
interaction. We therefore construct a nonlocal Gaussian potential in
the isospin basis by applying the Gauss-Legendre integration formula
to special functions appearing in the exchange RGM kernel.\cite{Fuk09}
The nonlocality and the energy dependence of the QM BB interaction 
is strictly preserved in the nonlocal Gaussian potential. We find that
the 15-point Gauss-Legendre integration formula is good enough
to carry out the few-body calculations accurately. The $NN$ phase shifts
predicted by this potential are essentially the same as those by
fss2 with accuracy of less than $0.1^{\circ}$.
We will show that the difference in the triton binding energy between
fss2 and this nonlocal Gaussian potential is only 15 keV.

Here, we study the low-energy neutron-deuteron ($nd$) elastic scattering 
below the deuteron breakup threshold based on the formulation developed
in Ref.\citen{ndsc1}. For this purpose, the $S$-wave
effective range theory in the channel-spin formalism is very useful.
The channel spin $S_{c}$ is composed of the $NN$ total angular momentum
$I$ and the spin 1/2 of the third nucleon.\cite{Sey69} Since the deuteron
channel with $I=1$ only survives in the asymptotic region, the scattering
amplitudes for the $nd$ elastic scattering are specified by the channel
spin $S_{c}=1/2$ (the spin-doublet channel) and $S_{c}=3/2$
(the spin-quartet channel). These two channels have quite different
characteristics with respect to the deuteron distortion effect. Namely,
in the spin-quartet channel the incident neutron can not penetrate deep
inside of the deuteron due to the effect of the Pauli principle, resulting
in the weak distortion effect of the deuteron. On the other hand, the
neutron can freely approach to the deuteron in the spin-doublet channel,
causing strong distortion effects reflecting strong sensitivity to
details of the $NN$ interaction. In this sense, the spin-doublet
scattering length $^{2}a_{nd}$ becomes an important observable to measure
if the $NN$ interaction is appropriate or not. It is known for a long
time that a larger triton binding energy corresponds to a smaller
$^{2}a_{nd}$. This linear correlation is known as the Phillips line,
\cite{Phi77} and is confirmed by many theoretical calculations.
\cite{Fri84,Chen86,Chen89,Chen91,Kiev94,Kiev97,Wit03,Ga07}
Since the binding energy of the triton is not reproduced in the $NN$ 
meson-exchange potentials, the experimental value, $^{2}a_{nd}=0.65
\pm 0.04$ fm, \cite{Di72} is not reproduced either. The calculated
$^{2}a_{nd}$ is more than 0.9 fm if only an $NN$ force is used.
It is therefore a common practice to add the $3N$ force to reproduce
the triton binding energy as well as the correct $^{2}a_{nd}$. A thorough
investigation of the triton binding energy and the scattering lengths,
using a number of meson-exchange potentials and various $3N$ forces,
is given in Ref.\,\citen{Wit03}. We can expect that the nonlocal effect
of fss2 leads to a good reproduction of the doublet scattering length
since fss2 gives a large triton binding energy close to the experiment.
In Ref.\,\citen{Ga07}, an application of the nonlocal interaction to the
$nd$ scattering lengths was made, using the $NN$ interaction based on the
chiral constituent quark model.\cite{Val00,Val05} Since their Faddeev
calculation does not treat the energy dependence of the QM $NN$ interaction
properly, they have obtained an insufficient triton binding energy
\cite{Ju02} and  a large doublet scattering length $^{2}a_{nd}$,\cite{Ga07}
almost comparable to the meson-exchange predictions.

The most accurate method to determine the scattering lengths is
to calculate the zero-energy scattering amplitude directly,
as carried out in Refs.\,\citen{Fri84,Chen86,Wit03,Ga07}, etc. In this approach, however,
the zero-energy $nd$ scattering is only examined. More extensive study
of the low-energy $nd$ elastic scattering can be achieved in terms
of the effective range theory,\cite{Phi77,Chen89} in which a pole
structure existing in the effective-range function for the doublet-$S$
channel should be properly taken into account.\cite{Del60,Phi69,Rei69,Whi76}
We can discuss the energy dependence of the low-energy $S$-wave phase
shifts in this approach. On the other hand, this method has a problem
of numerical inaccuracy at extremely low energies below 100 keV
in the center-of-mass (cm) system, since the solution of the basic
equation becomes very singular in the momentum representation.

In this paper, we start with the nonlocal Gaussian QM $NN$ interaction and
eliminate the energy dependence numerically in the above mentioned
renormalized RGM formalism. We then apply this interaction to the $nd$
scattering and solve the Alt-Grassberger-Sahndhas (AGS) equation \cite{AGS}
to obtain the scattering amplitudes. The elastic scattering amplitudes
are conveniently parameterized by the standard eigenphase shifts and
mixing parameters defined in Ref.\,\citen{Sey69}. The spin-doublet and
quartet $S$-wave effective range parameters: i.e., $^{2}a_{nd},
(\hbox{}^{2}{r_{e}})_{nd}, \hbox{}^{4}a_{nd}$ and $(^{4}r_{e})_{nd}$,
together with the pole parameter $q_{Q}$ in the doublet case, are
calculated by employing the $S$-wave single-channel effective range formula.
We find reasonable agreement of the low-energy differential cross sections
with the $nd$ experimental data. Since the $nd$ data have rather large
error bars, we will also evaluate the differential cross sections of the
$pd$ elastic scattering by employing a simple prescription, called the
Coulomb externally corrected approximation,\cite{Dol82,De05} to incorporate
the Coulomb effect with some modifications to the nuclear phase shifts.
We will find that the $^{2}S_{1/2}$ eigenphase shift predicted by our model
is sufficiently attractive to reproduce the doublet scattering length
$^{2}a_{nd}$ and the low-energy $nd$ and $pd$ differential cross
sections. It is possible that the correct treatment of the Coulomb
effect can reproduce the $pd$ differential cross sections and the
polarization observables below the deuteron breakup threshold 
without introducing the $3N$ force.

The organization of this paper is as follows. In \S \,2.1, a brief
description of the Faddeev formalism is given for the bound-state and
$nd$ scattering problems. The spin-isospin factors and rearrangement
factors for the permutation operator are explicitly given in Appendix A.
In \S \,2.2, we recapitulate the procedure to obtain the eigenphase shifts
and their $J$-averaged central phase shifts from the solutions of the
AGS equation. The single-channel effective-range expansion is explained
in \S \,2.3. The eigenphase shifts of our model are compared in \S \,3.1
with those of Argonne V18 (AV18) plus Urbana $3N$ force,
obtained by the $K$-harmonics technique.
The $nd$ and $pd$ differential cross sections below the deuteron breakup
threshold are also discussed. The effective range parameters
are given in \S \,3.2, together with the analysis of the $S$-wave
contributions to the $nd$ total cross sections. The last section is devoted
to a summary.

\section{Formulation}
\subsection{Faddeev approach to the triton and the $nd$ scattering}
We start with the three-body Schr\"{o}dinger equation
\begin{equation}
[E-H_{0}-V^{\rm RGM}_{\alpha}-V^{\rm RGM}_{\beta}-V^{\rm RGM}_{\gamma}]
\,\Psi=0~, \label{fad1}
\end{equation}
where $V^{\rm RGM}_{\alpha}$ denotes the energy-independent renormalized
RGM kernel for which the detailed derivation is given in Ref.\,\citen{ndsc1}.
The subscripts $ \alpha $, $ \beta $ and $ \gamma $ in Eq.\,(\ref{fad1})
specify the types of Jacobi coordinates related to the residual pair in
the usual way, with $(\alpha, \beta, \gamma)$ being the cyclic permutation
of (123). For the systems of three identical particles, the Faddeev
equation for the bound state reads
\begin{equation}
\psi=G_{0}tP \psi~, \label{fad2}
\end{equation}
where $P=P_{(12)}P_{(13)}+P_{(13)}P_{(12)}$ is a sum of the permutation
operators for the nucleon rearrangement.\cite{GLtext} The $NN$ $t$-matrix
in the three-body model space is derived from the standard
Lippmann-Schwinger equation $t=v+vG_{0}t$ with $v=V^{\rm RGM}$, where
$G_{0}=1/(E-H_{0})$ is the three-body Green function for the free motion.
It is composed of the negative total energy $E$ in the cm system and
the three-body kinetic-energy operator $H_{0}=h_{0}+\bar{h}_{0}$.
The operators $h_{0}$ and $ \bar{h}_{0}$ correspond to the kinetic energy
for ${\bm p}_{3}=(1/2)({\bm k}_{1}-{\bm k}_{2})$
and ${\bm q}_{3}=(1/3)(2{\bm k}_{3}-({\bm k}_{1}+{\bm k}_{2}))$ respectively
when the Jacobi coordinates with $ \gamma=3$ is chosen.
The vector ${\bm k}_{i}$ is the individual momentum of particle $i$.
The Faddeev component
$ \psi $ is defined through $ \psi=G_{0}v \Psi $, with the total wave
function $ \Psi =\sum _{\alpha}\psi _{\alpha}$ in Eq.\,(\ref{fad1}).
In the partial wave expansion, we use the channel-spin formalism specified
by $(I{\scriptsize{\frac{1}{2}}})S_{c}$. The relative angular momentum
$ \ell $ between the spectator nucleon and the $NN$ subsystem is coupled
with channel spin $S_{c}$ and makes the total angular momentum
$(\ell S_{c})J$. The $NN$ channel is specified by $(\lambda s)I;t$, where
$ \lambda,~s$ and $t$ are the orbital angular momentum, the spin and
the isospin of the $NN$ system, respectively. The $NN$ isospin $t$ is
uniquely specified by $ \lambda $ and $s$ from the Pauli principle
$(-)^{\lambda+s+t}=-1$. We further set the parity restriction
$ \pi=(-)^{\lambda+\ell}$, which is conserved for each $J$.
The angular-spin-isospin wave functions are thus defined by
\begin{eqnarray}
& & |{\bm p},{\bm q};123 \rangle=\sum _{\gamma}|p,q,\gamma \rangle
\langle \gamma|\widehat{\bm p},\widehat{\bm q};123 \rangle \nonumber~, \\
& & \langle \widehat{\bm p},\widehat{\bm q};123|{\gamma}\rangle=
[Y_{\ell}(\widehat{\bm q})[[Y_{\lambda}(\widehat{\bm p})\chi _{st}(1,2)]
_{I}\chi _{{\scriptsize{\frac{1}{2}}}{\scriptsize{\frac{1}{2}}}}(3)]
_{S_{c}}]_{JJ_{z};{\scriptsize{\frac{1}{2}}}T_{z}}~,
\label{LS-couple}
\end{eqnarray}
with $ \gamma=[\ell[(\lambda s)I{\scriptstyle{\frac{1}{2}}}]S_{c}]JJ_{z};
(t {\scriptstyle{\frac{1}{2}}}){\scriptstyle{\frac{1}{2}}} T_{z}$
in the channel-spin representation. The partial wave expansion of
the Faddeev equation in Eq.\,(\ref{fad2}) is given by
\begin{eqnarray}
& & \psi _{\gamma}(p,q)=-\frac{M}{\hbar^{2}}
\frac{1}{\kappa^{2}_{t}+p^{2}+(3/4)q^{2}}\sum _{\gamma',\gamma''}
\int^{\infty}_{0}~dq'q'^{2} \int^{\infty}_{0}~dp'p'^{2}
\int^{\infty}_{0}~dq''q''^{2} \nonumber \\
& & \times \int^{\infty}_{0}~dp''p''^{2}
\langle p,q,\gamma|t|p',q',\gamma' \rangle
\langle p',q',\gamma'|P|p'',q'',\gamma'' \rangle \psi _{\gamma''}(p'',q''),
\label{triton}
\end{eqnarray}
where $M=(M_{n}+M_{p})/2$ is the averaged nucleon mass in the isospin 
formalism. The binding energy of the triton is deduced from
$E_{B}=(-E)=(\hbar^{2}/M)\kappa^{2}_{t}$. The coupled integral
equation for $p$ and $q$ in Eq.\,(\ref{triton}) is solved in the
Lanczos-Arnoldi method after the necessary process of discretization. See
Ref.\,\citen{triton1} for details. The $NN$ $t$-matrix is factorized as
\begin{equation}
\langle p,q,\gamma|t|p',q',\gamma' \rangle
=\frac{4 \pi}{(2 \pi)^{3}}\frac{\delta(q-q')}{qq'}
t_{\gamma \gamma'}\left(p,p';E-\frac{3 \hbar^{2}}{4M}q'^{2}\right)~.
\label{fad5}
\end{equation}
The matrix element of the permutation operator is evaluated as
\begin{equation}
\langle p,q,\gamma|P|p',q',\gamma' \rangle=\frac{1}{2}\int^{1}_{-1}~dx
\frac{\delta(p-p_{1})}{p^{\lambda+2}}g_{\gamma,\gamma'}(q,q',x)
\frac{\delta(p'-p_{2})}{p'^{\lambda'+2}},~
\label{fad6}
\end{equation}
with $p_{1}=p(q',q/2;x)$, $p_{2}=p(q,q'/2;x)$ and $p(a,b;x)
=\sqrt{a^{2}+b^{2}+2abx}$. The basic rearrangement coefficients
$g_{\gamma,\gamma'}(q,q',x)$ contain the spin-isospin factors and the
explicit expression depends on a specific type of the 
channel-coupling scheme, as given in Appendix A.

For the $nd$ scattering, the three-body scattering amplitudes are obtained
by solving the AGS equation \cite{PREP}
\begin{equation}
U|\phi \rangle =G_{0}^{-1}P| \phi \rangle +PtG_{0}U|\phi \rangle~,
\label{AGS-1}
\end{equation}
where $|\phi \rangle =|{\bm q}_{0},\psi _{d} \rangle $ is the plane-wave
channel wave function with $|\psi _{d}\rangle $ being the deuteron wave
function. The total cm energy $E$ in ${G_{0}}^{-1}=E+i0-H_{0}$ is
expressed as $E=E_{\rm{cm}}+\varepsilon _{d}$, where $E_{\rm{cm}}=
(3 \hbar^{2}/4M)q^{2}_{0}$ is the cm incident energy of the neutron
and $|\varepsilon _{d}|=-\varepsilon _{d}$ is the deuteron binding energy.
In the AGS equation, we have in general two types of singularities, but the
notorious moving singularity does not appear for the energies below the
deuteron breakup threshold. The other singularity related to the deuteron
pole of the $NN$ $t$-matrix is directly incorporated to the AGS equation
in the Noyes-Kowalski method. For the detailed procedure to overcome
difficulties of these singularities, Ref.\,\citen{ndsc1} should be referred
to. Here, we recapitulate the method to derive the elastic scattering
amplitudes. The singularity of the $NN$ $t$-matrix is separated as
\begin{equation}
t=\tilde{t}-icG^{-1}_{0}| \phi \rangle \langle \phi|G^{-1}_{0}\quad
\hbox{with}\quad c=2 \pi \frac{q_{0}M}{3 \hbar^{2}}~.
\label{singulart}
\end{equation}
If we use the principal-value $t$-matrix $ \tilde{t}$ 
in Eq.\,(\ref{AGS-1}), we obtain
\begin{equation}
U| \phi \rangle=G^{-1}_{0}P|\phi \rangle[1-ic \langle \phi|U|\phi \rangle]
+P \tilde{t}G_{0}U|\phi \rangle ~.
\label{AGS-2}
\end{equation}
We eliminate the first term of Eq.\,(\ref{AGS-2}) by defining $W$ as 
\begin{equation}
W=G_{0}P-P|\phi \rangle Z^{-1}\langle \phi|P \quad \hbox{with}\quad
Z=\langle \phi |G^{-1}_{0}P|\phi \rangle~.
\label{fad10}
\end{equation}
Our basic equation is
\begin{equation}
\tilde{Q}|\phi \rangle=\tilde P|\phi \rangle 
+W \tilde{t}\tilde{Q}|\phi \rangle~, 
\label{fad11}
\end{equation}
where $ \tilde{Q}|\phi \rangle $ and $ \tilde{P}|\phi \rangle $ is 
defined through 
\begin{equation}
G_{0}U|\phi \rangle =\tilde{Q}|\phi \rangle \langle \phi|U|\phi \rangle,\quad
\tilde{P}|\phi \rangle=P|\phi \rangle Z^{-1}~.
\label{fad12}
\end{equation}
The elastic scattering amplitudes are obtained by multiplying
Eq.\,(\ref{AGS-2}) with $Z^{-1}\langle \phi|$ from the left-hand side.
One can easily show
\begin{equation}
\langle \phi|U|\phi \rangle =[Z^{-1}-\langle \phi|X|\phi \rangle
+ic]^{-1}\quad \hbox{with} \quad \langle \phi|X|\phi \rangle
=\langle \phi|\tilde{P}\tilde{t}\tilde{Q}|\phi \rangle~.
\label{fad13}
\end{equation}
The partial wave components of the scattering amplitude,
$U^{J}_{(\ell'S'_{c}),(\ell S_{c})}=\langle 
\phi _{\ell' S'_{c}}|U|\phi _{\ell S_{c}} \rangle $, are defined by
\begin{eqnarray}
& & \langle \phi _{{\bm q}_{f}};S'_{c}S'_{cz}|U
|\phi _{{\bm q}_{i}};S_{c}S_{cz} \rangle 
=\sum _{\ell'\ell JJ_{z}}U^{J}_{(\ell'S'_{c}),(\ell S_{c})}  \nonumber \\
&\times& \sum _{m'}\langle \ell'm'
S'_{c}S'_{cz}|JJ_{z} \rangle 
Y_{\ell'm'}({\widehat{\bm q}_{f}})
\sum _{m}\langle \ell mS_{c}S_{cz}|JJ_{z} \rangle 
Y^{*}_{\ell m}({\widehat{\bm q}_{i}})~,
\label{fad14}
\end{eqnarray}
with $|{\bm q}_{f}|=|{\bm q}_{i}|=q_{0}$.
Once the partial wave components of $ \langle \phi|X|\phi \rangle $
is calculated, $U^{J}_{(\ell' S'_{c}),(\ell S_{c})}$ is obtained by
solving an equation,
\begin{equation}
\sum_{\ell',S'_{c}}\left[(Z^{-1})_{\ell S_{c},\ell'S'_{c}}
-X_{\ell S_{c},\ell'S'_{c}}+ic~\delta _{\ell,\ell'}\delta _{S_{c},S'_{c}}
\right]~U_{(\ell'S'_{c}),(\ell''S''_{c})}
=\delta _{\ell,\ell''}\delta _{S_{c},S''_{c}}~.
\label{fad15}
\end{equation}
The coupled channel $S$-matrix is given by
$S^{J}_{(\ell S_{c}),(\ell'S'_{c})}=\delta _{\ell,\ell'}\delta _{S_{c},S'_{c}}
-2ic~U^{J}_{(\ell S_{c}),(\ell'S'_{c})}$. 

\subsection{Eigenphase shifts}
In the channel-spin representation, the asymptotic channel wave function
\linebreak $|\phi;(\ell S_{c})JJ_{z}\rangle $ in the partial wave expansion
is specified by $(\ell S_{c})J=~(J \pm 3/2,3/2)J,$ \linebreak $(J \mp 1/2,1/2)J,$ 
 and $(J \mp 1/2,3/2)J$ for the parity $ \pi=(-)^{J \mp 1/2}$.
Therefore, the $S$-matrix $S^{J}_{(\ell S_{c}),(\ell' S'_{c})}$ is 
a two-dimensional matrix for $J=1/2$, and a three-dimensional matrix
for $J \neq 1/2$. The $S$-matrix can be diagonalized as
\begin{equation}
S=U^{\dagger}e^{2i \Delta}U~, \label{fad16}
\end{equation}
where $ \Delta $ is the diagonal matrix of the eigenphase shifts
$ \delta^{J}_{\ell S_{c}}$. (We follow the notation in Ref.\,\citen{Sey69},
but we should note that $ \ell $ and $S_{c}$ are not good quantum numbers.)
The matrix $U$ can be parameterized in terms of the mixing parameters
$ \varepsilon, \xi $ and $ \eta $,\cite{Sey69}
\begin{eqnarray}
& & U=\begin{pmatrix}
1 & 0 & 0 \\
0 & \cos{\varepsilon} & \sin{\varepsilon} \\
0 & -\sin{\varepsilon} & \cos{\varepsilon} \\
\end{pmatrix}
\begin{pmatrix}
\cos{\xi} & 0 & \sin{\xi} \\
0 & 1 & 0 \\
-\sin{\xi} & 0 & \cos{\xi} \\
\end{pmatrix}
\begin{pmatrix}
\cos{\eta} & \sin{\eta} & 0 \\
-\sin{\eta} & \cos{\eta} & 0 \\
0 & 0 & 1 \\
\end{pmatrix}
\nonumber \\
& & =\begin{pmatrix}
\cos{\xi}\cos{\eta} & \cos{\xi}\sin{\eta} & \sin{\xi} \\
-\cos{\varepsilon}\sin{\eta}-\sin{\varepsilon}\sin{\xi}\cos{\eta} 
& \cos{\varepsilon}\cos{\eta}-\sin{\varepsilon}\sin{\xi}\sin{\eta}
& \sin{\varepsilon}\cos{\xi} \\
\sin{\varepsilon}\sin{\eta}-\cos{\varepsilon}\sin{\xi}\cos{\eta} 
& -\sin{\varepsilon}\cos{\eta}-\cos{\varepsilon}\sin{\xi}\sin{\eta}
& \cos{\varepsilon}\cos{\xi}
\end{pmatrix}~. \nonumber \\ \label{eig2}
\end{eqnarray}
We assume $ \varepsilon =\xi=0$ for the $J^{\pi}=1/2^{+}$ and
$ \xi=\eta=0$ for the $J^{\pi}=1/2^{-}$.
By using a function,
\begin{equation}
{\rm Arctan}~{z}=\frac{1}{2i}{\rm Log}{\frac{1+iz}{1-iz}}\ , \label{fad18}
\end{equation}
we can calculate the mixing parameters $ \varepsilon, \xi $ and $ \eta $
from the following equations:
\begin{eqnarray}
\eta &=& {\rm Arctan}{\frac{U_{12}}{U_{11}}},\quad 
\xi=-{\rm Arctan}{\left(\frac{U_{31}\cos{\eta}
+U_{32}\sin{\eta}}{U_{33}}\right)},\quad \nonumber \\
\varepsilon &=& -{\rm Arctan}{\left(\frac{U_{31}\sin{\eta}-U_{32}\cos{\eta}}
{U_{21}\sin{\eta}-U_{22}\cos{\eta}}\right)}~,
\label{eig3}
\end{eqnarray}
where $U_{ij}$ is the $(i,j)$ component of the matrix $U$.

Before applying Eq.\,(\ref{eig3}), we should reorder the eigenstates
of Eq.\,(\ref{fad16}) such that the leading term of the eigenvectors
follows the decreasing order. This prescription corresponds to the rule
to assign the eigenphases to such quantum numbers as having the
dominant components of the eigenvectors.\footnote{
This rule is not applied to the case of $^{2}P_{3/2}$--$^{4}P_{3/2}$ coupling for
$E_{\rm cm}>40$ MeV, where the magnitude of the mixing parameter
$ \varepsilon $ exceeds $45^{\circ}$.}
For energies above the breakup
threshold, all the angles become complex. The largest component of
the eigenvector is given by a real positive number. This gives an
ambiguity of the sign of the mixing angles, which in turn comes from
the phase convention of the eigenstates. We should note that Eq.\,(\ref{eig3})
is most convenient if the diagonal $U_{ii}$ components are close to one.
This is because the mixing angles are usually very small with the magnitude of
several degrees (except for $E_{\rm cm} >$ a few MeV in the
$^{2}{P}_{J}$--${}^{4}P_{J}$ couplings). 

Since the difference in the eigenphase shifts with the same $(\ell S_{c})$ 
but different $J$ is very small, the $J$-averaged central phase shift
$ \delta^{C}_{\ell S_{c}}$ is convenient to discuss the scattering cross
sections approximately. The $J$-averaged central phase shift is defined by
taking an average of all possible $J$ states $^{2S_{c}+1}\ell _{J}$ with 
the weight factor $(2J+1)$,
\begin{equation}
\delta^{C}_{\ell S_{c}}= \frac{1}{(2 \ell+1)(2S_{c}+1)}
\sum _{J}(2J+1)\,\delta^{J}_{\ell S_{c}}~.
\label{C-phase}
\end{equation}
The differential cross section for the $nd$ elastic scattering is
calculated by summing over the final spin states and by taking an average
over the initial spin states. This is given by
\begin{equation}
\frac{d \sigma}{d \Omega}=\frac{1}{3}
\left(\frac{d \sigma}{d \Omega}\right)^{S_{c}=1/2}
+\frac{2}{3}\left(\frac{d \sigma}{d \Omega}\right)^{S_{c}=3/2}~,
\label{DCS-1}
\end{equation}
where the differential cross section for each channel spin $S_{c}$
is calculated from the scattering amplitudes
$f_{\ell}=(1/q_{0})e^{i \delta _{\ell}}\sin{\delta _{\ell}}$ through
\begin{eqnarray}
\left(\frac{d \sigma}{d \Omega}\right)^{S_{c}} &=& \left|\sum _{\ell}
(2 \ell+1)\,f_{\ell}\,P_{\ell}(\cos{\theta})\right|^{2}~.
\label{DCS-2}
\end{eqnarray}
Here, we abbreviate the $J$-averaged central phase shift
$ \delta^{C}_{\ell S_{c}}$ to $ \delta _{\ell}$ for each $S_{c}$.
In order to compare our results with the experimental data,
we have to take into account the Coulomb force for the $pd$ scattering.
The comparison with the $pd$ data is desirable since they are abundant
and more accurate than those of the $nd$ scattering.
For example, in Refs.\,\citen{Sc72,Arv74,Ch75,Kn93,Kiev96,To02} many 
phase shift analyses have been carried out with high accuracy.
The exact treatment of the Coulomb force in the three-body
problem is still a challenging task.\cite{Dol82,De05,Alt78,Ber90}
Because of its long-range nature, the Coulomb potential is not amenable to
the standard scattering theory. We therefore incorporate the Coulomb
effect to evaluate the differential cross sections of the $pd$ elastic
scattering, by employing a following simple prescription with some
modifications to the nuclear phase shifts. The single channel differential
cross section is given by,
\begin{eqnarray}
\left(\frac{d \sigma}{d \Omega}\right)^{S_{c}} &=& \left
|e^{-2i \sigma _{0}}f^{C}(\theta)+\sum _{\ell}(2 \ell+1)
e^{2i(\sigma _{\ell}-\sigma _{0})}f^N_{\ell}
P_{\ell}(\cos{\theta})\right|^{2}~,
\label{DCS-3}
\end{eqnarray}
where $f^{C}(\theta)$ is the standard Coulomb scattering amplitude.
The scattering amplitude from the strong interaction, $f^{N}_{\ell}$
in Eq.\,(\ref{DCS-3}), is estimated from the $nd$ eigenphase shifts of
fss2 by adding the difference of the corresponding quantities for the
$pd$ and $nd$ scattering, which will be taken from published results
for other $NN$ interactions. In Eq.\,(\ref{DCS-3}), the partial-wave
Coulomb phase shift $ \sigma _{\ell}$ is given by $ \sigma _{\ell}
={\rm arg}\,{\rm \Gamma}(\ell+1+i \eta)$ with the Sommerfeld parameter
$ \eta $ for the relative motion of the proton and the deuteron.

\subsection{Effective-range expansion for the $nd$ scattering}
If we use the effective range theory, we can study energy dependence
of the phase shifts reflected in the effective range $r_{e}$
and discuss contributions to the total cross sections from the
$S$-wave components between the neutron and the deuteron.
We first calculate the eigenphase shifts for $J^{\pi}=1/2^{+}$ and
$3/2^{+}$ states. In the case of $J^{\pi}=1/2^{+}$, the $^{2}S_{1/2}$ 
and $^{4}D_{1/2}$ channels are coupled. For the neutron energies
below the deuteron breakup threshold, the phase shifts and
mixing parameters are small except for the dominant $S$-wave 
eigenphase shift, so that the effective-range expansion formula for a
single channel problem can be safely applied to this component to
obtain the effective range parameters. This is also the case for the
$J^{\pi}=3/2^{+}$ state, where the $^{4}S_{3/2}$, $^{2}D_{3/2}$ and
$^{4}D_{3/2}$ channels are coupled. In the quartet $S$-channel,
we can expand the effective-range function $K(q_{0})=q_{0}\cot{\delta}$
in the power series of $q^{2}_{0}$:
\begin{equation}
K(q_{0})=-\frac{1}{a}+\frac{1}{2}{r_e}q^{2}_{0}+{\cal O}(q^{4}_{0})~,
\label{quartet}
\end{equation}
where $a$ is the scattering length, $r_{e}$ is the effective range, and
$ \delta=\delta^{{\scriptstyle \frac{3}{2}}}_{0{{\scriptstyle \frac{3}{2}}}}$
in $K(q_{0})=q_{0}\cot{\delta}$ is the $S$-wave eigenphase shift
for the quartet channel with $q_{0}$ being the relative wave number between
the neutron and the deuteron.

In the doublet-$S$ channel, the effective-range function $K(q_{0})$,
which is the real part of the inverse scattering amplitude, has a
pole just below the elastic threshold. \cite{Del60,Rei69,Phi69,Whi76}.
We parameterize the effective-range function in the doublet channel as
\begin{equation}
K(q_{0})=\frac{-\frac{1}{a}+\frac{1}{2}r_{e}q^{2}_{0}+{\cal O}(q^{4}_{0})}
{1+\left(q_{0}/q_{Q}\right)^{2}}~,  \label{doublet}
\end{equation}
where a pole parameter $q_{Q}$ specifies the pole position, and
$ \delta=\delta^{{\scriptstyle \frac{1}{2}}}_{0{{\scriptstyle \frac{1}{2}}}}$
is the $S$-wave eigenphase shift for the doublet channel.
The origin of this pole structure is studied by means of $N/D$ equations 
.\cite{Rei69,Phi69,Whi76}. In the $N/D$ formalism, the partial wave components of
the scattering amplitude are given by ${N(z)}/{D(z)}$, where $N(z)$ and $D(z)$
are the analytic functions of the complex and dimensionless energy variable $z$ defined by
$z=E_{\rm cm}/|\varepsilon _{d}|$. The two and three-body unitarities and
the cut structure in the negative energy, which are shown in Fig.\,1, yield
the relationship, $N(z)=-z^{-\frac{1}{2}}{\rm Im}~D(z)$.
The $N/D$ equations are usually constructed by applying the Kramers-Kronig
relations to $N(z)$ and $D(z)$.
From the solution of the $N/D$ equation, it was found that this singularity are
brought about by both the dominant single-nucleon exchange and the other effects
such as the two-nucleon exchange etc. slowly varying for small $z$.
In the doublet channel, the single-nucleon exchange, which is
by far the longest-range force, is attractive. Thus, nothing prevents the
other effects from influencing the low-energy scattering. On the other hand, 
the single-nucleon exchange is strongly repulsive in the quartet channel.
Therefore, the nucleons in this channel can not penetrate to the region where the
other attractive forces could act.  This is why the pole structure is
found only in the doublet channel.
\begin{center}
\begin{figure}[t]
\centerline{\includegraphics[scale=0.5,
trim= 0 170 0 130]{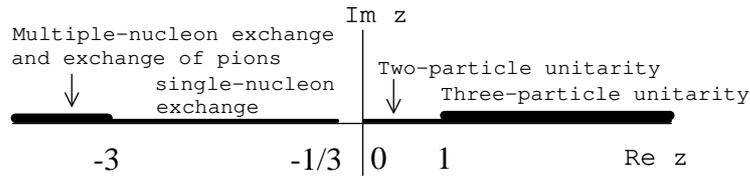}}
\caption{The cut structure used in the $N/D$ formalism
for the $nd$ elastic scattering amplitude
in the complex $z$ plane.\cite{Whi76} Here, $z$ is defined by
$z={E_{\rm cm}}/{|\varepsilon _{d}|}$.} \label{fig1}
\vspace{-4mm}
\end{figure}
\end{center}

\section{Results and Discussion}
\subsection{Eigenphase shifts}
\begin{figure}[htbp]
\centerline{\includegraphics[scale=0.4, angle=0,
trim=10 10 10 170]{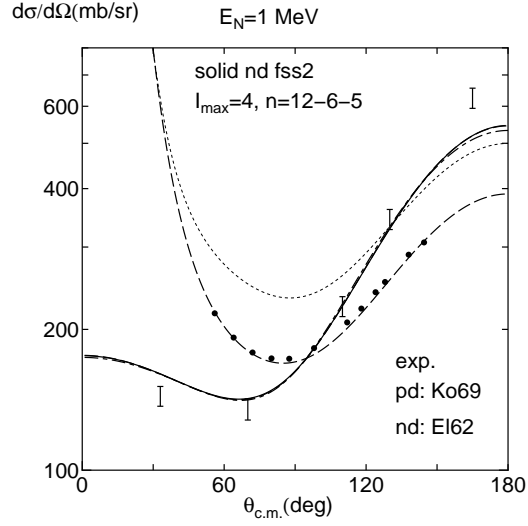}}
\vspace{-20mm}
\caption{fss2 predictions to the $nd$ and $pd$ differential cross
sections obtained from various prescriptions for the phase shifts at
$E_{N}$=1 MeV: the exact $nd$ calculation (solid curve), the $J$-averaged
central phase shifts (dot-dashed curve), the Coulomb externally corrected
approximation with $ \delta^{N}=\delta(nd)$ (dotted curve), and the
Coulomb modified nuclear phase shifts in Table III (dashed curve). 
The experimental data are taken from Ref.\,\citen{El62} for El62
($nd$ with errorbars) and Ref.\,\citen{Ko69} for Ko69 ($pd$ with filled circles).}
\label{FDCS-1}
\end{figure}

The direct comparison of the $nd$ eigenphase shifts, predicted by fss2,
with the results of the modern phase shift analysis for the $pd$ scattering
is not possible because of the Coulomb effect. We therefore list in Table
\ref{table1} our results for the energies of $E_{n}$=1, 2 and 3 MeV
together with other theoretical predictions by the Pisa group,\cite{Kiev96}
which are calculated using Argonne V18 $NN$ potential (AV18)\cite{AV18}
and AV18+Urbana(UR) $3N$ potentials.\cite{UR3N} Here, $E_{n}=(3/2)E_{\rm cm}$
is the neutron incident energy measured in the laboratory system.
We have included the $NN$ interaction up to the total angular momentum
$I_{\rm max}$=4 and the momentum mesh points $n \equiv n_{1}$-$n_{2}$-$n_{3}$
=12-6-5 in the notation defined in \S 3.1 of Ref.\,\citen{ndsc1}.
The UR $3N$ potential gives a sizable effect
of about three to four degrees only on the $J^{\pi}=1/2^{+}$ channel. We
immediately find an outstanding feature in the $J^{\pi}=1/2^{+}$ channel. Namely,
our results by fss2 are very similar not to the AV18 results but to the
AV18+UR($3N$) results shown in the parentheses. It is not surprising since
fss2 gives the nearly correct binding energy without introducing the $3N$ force.
On the other hand, the phase shift parameters of the $J^{\pi}=3/2^{+}$ state are
very similar between fss2 and AV18. In this state, the effect of the UR $3N$
force is very small owing to the Pauli principle. The difference between fss2
and AV18 is less than $0.2^{\circ}$ -- $0.3^{\circ}$, which is comparable to
the effect of the $3N$ force. 
For the $P$ states, some of
the eigenphase shifts show somewhat larger difference from the AV18 and
AV18+UR($3N$) results especially at $E_{n}=3$ MeV,
but still less than $1^{\circ}$ difference. After all, we have found
good correspondence between our fss2 results and the predictions
by the AV18+UR($3N$) potentials. The resemblance seen in Table \ref{table1}
becomes more transparent if we calculate the $J$-averaged central phase
shifts defined by Eq.\,(\ref{C-phase}) and compare them. The $J$-averaged central
phase shifts can be used to evaluate the $nd$ differential cross sections
through Eqs.\,(\ref{DCS-1}) and ({\ref{DCS-2}}). We have illustrated these in
Figs.\,\ref{FDCS-1} -- 4 with dot-dashed curves, but they almost overlap
with the solid curves for the exact calculations. We find that the
$D$-wave components give an appreciable contribution to the differential
cross sections even in such a low energy as $E_{n}$=1 MeV. This is
of course because of the $D$-wave component of the deuteron wave function.
This analysis encourages us to study the $pd$ differential cross sections
by a simple approximation for the Coulomb effect, discussed in \S \,2.2.
We first assume that the nuclear scattering amplitude $f^{N}_{\ell}$ in
Eq.\,(\ref{DCS-3}) is equal to the $nd$ scattering amplitude from the
$J$-averaged central phase shift, 
$f_{\ell}=(e^{2i \delta^{C}_{\ell S_{c}}}-1)/2iq_{0}$.
This prescription yields fairly large overestimation of the differential 
cross sections as plotted in Figs.\,\ref{FDCS-1} -- 4 with dotted curves.
We find that a large effect of Coulomb modification on $f^{N}_{\ell}$
is necessary for the present low-energy $pd$ scattering, which is a
well-known fact claimed by many authors.\cite{Dol82,De05,Alt78,Ber90}
Here, we use an extended version of the Coulomb externally corrected 
approximation \cite{De05}, in which the nuclear eigenphase shifts for
the $pd$ scattering, $ \delta^{C}_{\ell S_{c}}(pd)$, are calculated 
from our $ \delta^{C}_{\ell S_{c}}(nd)$ by adding the difference of
those evaluated with another $NN$ interaction. 
\linebreak
\begin{table}[H]
\begin{center}
\caption{The $nd$ eigenphase shifts and mixing parameters (in degrees),
obtained from the model fss2. The maximum angular momentum for the $NN$
system, $I_{\rm max}=4$, and the momentum mesh points $n=12$-6-5
are used. The corresponding parameters calculated by the Pisa group
from the AV18 potential models are also listed for comparison.
\cite{Kiev96} The parameters in the parentheses are predictions by the
AV18+UR($3N$) potentials.}\label{table1} 
\setlength{\tabcolsep}{4.1mm}
\renewcommand{\baselinestretch}{0.95}
\vspace{0mm}
\begin{tabular}{ccccccc}
\hline \hline
Model & fss2 & AV18 & fss2 & AV18 & fss2 & AV18 \\ \hline
$E_{n}$(MeV) & 1.0 & 1.0 & 2.0 & 2.0 & 3.0 & 3.0 \\ \hline
$^{4}D_{1/2}$ & $-0.978$ & $-0.980$ & $-2.52$ & $-2.53$ & $-3.82$ & $-3.85$ \\
&---& ($-0.976$) &---& ($-2.52$) &---& ($-3.84$) \\
$^{2}S_{1/2}$ & $-14.8$ & $-18.1$ & $-24.2$ & $-28.3$ & $-30.8$ & $-35.3$ \\
&---& ($-14.3$) &---& ($-24.0$) &---& ($-30.8$) \\
$ \eta _{1/2+} $ & 1.29 & 0.928 & 1.48 & 1.08 & 1.55 & 1.12 \\
&---& (1.39) &---& (1.47) &---& (1.45) \vspace{2.5mm} \\ 
$^{2}P_{1/2}$ & $-4.12$ & $-4.13$ & $-6.55$ & $-6.57$ & $-7.43$ & $-7.49$ \\
&---& ($-4.13$) &---& ($-6.58$) &---& ($-7.50$) \\
$^{4}P_{1/2}$ & 11.8 & 12.0 & 19.3 & 19.9 & 23.6 & 24.2 \\ 
&---& (12.1) &---& (20.1) &---& (24.5) \\
$ \varepsilon _{1/2-} $ & 3.50 & 3.47 & 5.02 & 4.98 & 6.76 & 6.68 \\
&---& (3.53) &---& (5.07) &---& (6.82) \vspace{2.5mm}  \\
$^{4}S_{3/2}$ & $-46.6$ & $-46.7$ & $-60.5$ & $-60.8$ & $-69.6$ & $-69.9$ \\
&---& ($-46.6$) &---& ($-60.7$) &---& ($-69.7$) \\
$^{2}D_{3/2}$ & 0.564 & 0.564 & 1.50 & 1.51 & 2.34 & 2.36 \\
&---& (0.564) &---& (1.51) &---& (2.36) \\
$^{4}D_{3/2}$ & $-1.05$ & $-1.05$ & $-2.71$ & $-2.72$ & $-4.12$ & $-4.14$ \\
&---& ($-1.05$) &---& ($-2.71$) &---& ($-4.14$) \\
$ \varepsilon _{3/2+} $ & 0.603 & 0.621 & 0.688 & 0.686 & 0.763 & 0.747 \\
&---& (0.623) &---& (0.688) &---& (0.754) \\
$ \xi _{3/2+}$ & 0.516 & 0.511 & 0.957 & 0.948 & 1.36 & 1.35 \\
&---& (0.514) &---& (0.948) &---& (1.35) \\
$ \eta _{3/2+} $ & $-0.106$ & $-0.107$ & $-0.232$ & $-0.231$ & $-0.363$ & $-0.363$ \\
&---& ($-0.105$) &---& ($-0.228$) &---& ($-0.356$) \vspace{2.5mm} \\
$^{4}F_{3/2}$ & 0.122 & 0.121 & 0.491 & 0.488 & 0.919 & 0.920 \\
&---& (0.121) &---& (0.489) &---& (0.921) \\
$^{2}P_{3/2}$ & $-4.06$ & $-4.08$ & $-6.38$ & $-6.41$ & $-7.10$ & $-7.18$ \\ 
&---& ($-4.08$) &---& ($-6.43$) &---& ($-7.20$) \\
$^{4}P_{3/2}$ & 13.7 & 13.9 & 21.9 & 22.3 & 25.5 & 26.0 \\
&---& (14.0) &---& (22.3) &---& (26.0) \\
$ \varepsilon _{3/2-} $ & $-1.28$ & $-1.24$ & $-1.93$ & $-1.86$ & $-2.72$ & $-2.62$ \\
&---& ($-1.27$) &---& ($-1.89$) &---& ($-2.66$) \\
$ \xi _{3/2-}$ & $-0.196$ & $-0.177$ & $-0.334$ & $-0.262$ & $-0.427$ & $-0.265$ \\
&---& ($-0.177$) &---& ($-0.259$) &---& ($-0.256$) \\
$ \eta _{3/2-}$ & $-1.04$ & $-1.00$ & $-2.17$ & $-2.17$ & $-3.57$ & $-3.52$ \\
&---& ($-1.04$) &---& ($-2.16$) &---& ($-3.53$) \vspace{2.5mm} \\ 
$^{4}G_{5/2}$ & $-0.015$ & $-0.015$ & $-0.091$ & $-0.090$ & $-0.206$ & $-0.206$ \\
$^{2}D_{5/2}$ & 0.560 & 0.559 & 1.49 & 1.49 & 2.31 & 2.33 \\
$^{4}D_{5/2}$ & $-1.11$ & $-1.11$ & $-2.90$ & $-2.90$ & $-4.44$ & $-4.46$ \\
$ \varepsilon _{5/2+}$ & $-0.263$ & $-0.277$ & $-0.291$ & $-0.297$ & $-0.312$ & $-0.315$ \\
$ \xi _{5/2+}$ & $-0.233$ & $-0.272$ & $-0.491$ & $-0.494$ & $-0.716$ & $-0.701$ \\
$ \eta _{5/2+}$ & $-0.659$ & $-0.821$ & $-1.42$ & $-1.49$ & $-2.02$ & $-2.04$ \\
\hline
\end{tabular}
\end{center}
\end{table}

\addtocounter{table}{-1}
\begin{table}
\caption{--- continued}
\setlength{\tabcolsep}{4.4mm}
\renewcommand{\baselinestretch}{0.95}
\begin{center}
\begin{tabular}{ccccccc}
\hline \hline
Model & fss2 & AV18 & fss2 & AV18 & fss2 & AV18 \\ \hline
$E_{n}$(MeV) & 1.0 & 1.0 & 2.0 & 2.0 & 3.0 & 3.0 \\ \hline
$^{4}P_{5/2}$ & 13.1 & 13.2 & 21.4 & 21.7 & 25.8 & 26.0 \\ 
&---& (13.2) &---& (21.8) &---& (26.3) \\
$^{2}F_{5/2}$ & $-0.063$ & $-0.063$ & $-0.251$ & $-0.251$ & $-0.465$ & $-0.466$ \\
$^{4}F_{5/2}$ & 0.127 & 0.127 & 0.514 & 0.510 & 0.947 & 0.951\\
$ \varepsilon _{5/2-}$ & 0.399 & 0.447 & 0.479 & 0.472 & 0.514 & 0.538\\
$ \xi _{5/2-}$ & 0.384 & 0.390 & 0.690 & 0.684 & 0.938 & 0.926 \\
$ \eta _{5/2-}$ & $-0.117$ & $-0.123$ & $-0.239$ & $-0.239$ & $-0.343$ & $-0.334$
\vspace{3mm}\\
$^{4}D_{7/2}$ & $-1.03$ & $-1.03$ & $-2.66$ & $-2.67$ & $-4.04$ & $-4.06$ \\
$^{2}G_{7/2}$ & 0.0076 & 0.0075 & 0.047 & 0.047 & 0.108 & 0.107 \\
$^{4}G_{7/2}$ & $-0.015$ & $-0.015$ & $-0.094$ & $-0.095$ & $-0.215$ & $-0.214$\\
$ \varepsilon _{7/2+} $ & 0.230 & 0.325 & 0.353 & 0.368 & 0.355 & 0.355 \\
$ \xi _{7/2+}$ & 0.362 & 0.427 & 0.781 & 0.798 & 1.16 & 1.14 \\
$ \eta _{7/2+}$ & $-0.106$  & $-0.143$ & $-0.287$ & $-0.299$ & $-0.459$ & $-0.459$ 
\vspace{3mm}\\
$^{2}F_{7/2}$ & $-0.062$ & $-0.062$ & $-0.248$ & $-0.248$ & $-0.458$ & $-0.460$ \\
$^{4}F_{7/2}$ & 0.132 & 0.132 & 0.533 & 0.532 & 1.00 & 1.00 \\
$ \varepsilon _{7/2-}$ & $-0.207$ & 0.238 & $-0.244$ & $-0.236$ & $-0.256$ & $-0.232$ \vspace{3mm} \\
$^{2}G_{9/2}$ & 0.075 & 0.0075 & 0.047 & 0.046 & 0.106 & 0.105 \\
$^{4}G_{9/2}$ & $-0.015$ & $-0.016$ & $-0.097$ & $-0.097$ & $-0.223$ & $-0.223$ \\
$ \varepsilon _{9/2+}$ & $-0.125$ & $-0.183$ & $-0.189$ & $-0.199$ & $-0.173$ & $-0.176$ \vspace{3mm} \\
$^{4}F_{9/2}$ & 0.124 & 0.124 & 0.497 & 0.496 & 0.921 & 0.922 \vspace{3mm} \\
$^{4}G_{11/2}$ & $-0.015$ & $-0.015$ & $-0.092$ & $-0.091$ & $-0.208$ & $-0.208$ \\
\hline
\end{tabular}
\end{center}
\end{table}

\begin{table}[H]
\renewcommand{\baselinestretch}{1}
\setlength{\tabcolsep}{4.1mm}
\caption{The $nd$ $J$-averaged central phase shifts (in degrees),
calculated from the eigenphase shifts in Table I through
Eq.~(\ref{C-phase}). The others are the same as Table I.}
\begin{center}
\begin{tabular}{ccccccc}
\hline \hline
Model & fss2 & AV18 & fss2 & AV18 & fss2 & AV18 \\ \hline
$E_{n}$(MeV) & 1.0 & 1.0 & 2.0 & 2.0 & 3.0 & 3.0 \\ \hline
$^{2}S$ & $-14.8$ & $-18.1$  & $-24.2$ & $-28.3$ & $-30.8$ & $-35.3$ \\
&---& ($-14.3$) &--& ($-24.0$) &---& ($-30.8$) \\
$^{2}P$ & $-4.08$ & $-4.10$ & $-6.44$ & $-6.46$ & $-7.21$ & $-7.28$\\
&---& ($-4.10$) &---& ($-6.48$) &---& ($-7.30$)\\
$^{2}D$ & 0.562 & 0.562 & 1.49 & 1.50 & 2.32 & 2.34 \\
$^{2}F$ & $-0.0626$ & $-0.0624$ & $-0.249$ & $-0.249$ & $-0.462$ & $-0.463$ \vspace{3mm} \\
$^{4}S$ & $-46.6$ & $-46.7$ & $-60.5$ & $-60.8$ & $-69.6$ & $-69.9$ \\
&---& ($-46.6$) &---& ($-60.7$) &---& ($-69.7$) \\
$^{4}P$ & 13.1 & 13.2 & 21.2 & 21.6 & 25.3 & 25.7 \\
&---& (13.3) &---& (21.7) &---& (25.9) \\
$^{4}D$ & $-1.05$ & $-1.05$ & $-2.73$ & $-2.74$ &$-4.15$ & $-4.18$ \\
$^{4}F$ & 0.127 & 0.127 & 0.510 & 0.508 & 0.948 & 0.950 \\ \hline 
\label{table2}
\end{tabular}
\end{center}
\end{table}

\begin{table}[htbp]
\caption{The Coulomb modified $J$-averaged central phase shifts (in degrees),
$ \delta^{N}=\delta^{C}_{\ell S_{c}}(nd)+\Delta _{\ell}$, used for
the calculations of the $pd$ differential cross sections in Figs.~2 -- 4.
Here, $ \delta^{C}_{\ell S_{c}}(nd)$ are given in Table \ref{table2} and
$ \Delta _{\ell}=[\delta^{C}_{\ell S_{c}}(pd)-\delta^{C}_{\ell S_{c}}
(nd)]_{\rm AV18}$ are evaluated from Ref.\,\citen{Kiev96}. The results
of the phase shift analysis (PSA) at $E_{N}=3$ MeV are also shown for
comparison.}
\label{table3}
\begin{center}
\renewcommand{\baselinestretch}{1}
\begin{tabular}{ccccccccc}
\hline \hline
$E_{N}$(MeV) & \multicolumn{2}{c}{1.0} &  \multicolumn{2}{c}{2.0} 
& \multicolumn{2}{c}{3.0} & 3.0 (PSA) & 3.0 (PSA) \\ \hline
& $ \Delta _{\ell}$ & $ \delta^{N}$ & $ \Delta _{\ell}$ & $ \delta^{N}$ 
& $ \Delta _{\ell}$ & $ \delta^{N}$ & Ref.\,\protect \citen{Kiev96} 
& Ref.\,\protect\citen{Kn93} \\ \hline
$^{2}S$ & 4.9 & $-9.9$ & 4.1 & $-20.1$ & 3.1 & $-27.7$ 
& $-24.85 \pm 0.23$ & $-24.87 \pm 0.28$ \\
$^{4}S$ & 9.7 & $-36.9$ & 7.8 & $-52.7$ & 6.8 & $-62.8$ 
& $-63.80 \pm 0.11$ & $-63.95 \pm 0.28$ \\
$^{4}P$ & $-3.1$ & 10.0 & $-2.6$ & 18.6 & $-2.0$ & 23.3 
& $23.86 \pm 0.01$ & $23.37 \pm 0.11$ \\ \hline
\end{tabular}
\end{center}
\label{mod}
\end{table}

\begin{figure}[htbp]
\begin{minipage}{0.475\hsize}
\begin{center}
\includegraphics[width=\textwidth,
trim=50 30 50 180]{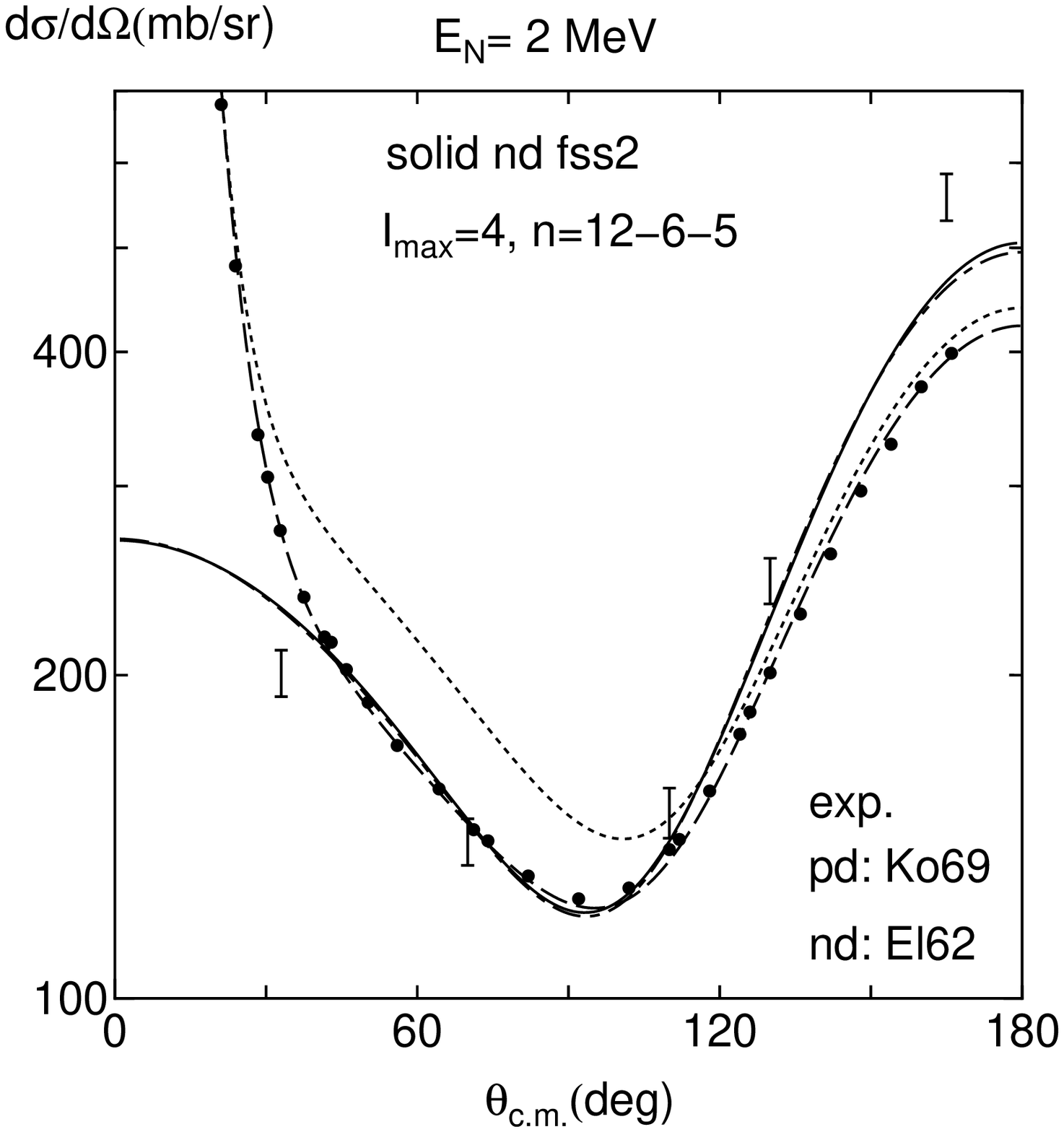}
\end{center}
\label{FDCS-2}
\vspace{-15mm}
\caption{The same as Fig.\,2, but for $E_{N}$=2 MeV.}
\end{minipage}
\hspace{5mm}
\begin{minipage}{0.475\hsize}
\begin{center}
\includegraphics[width=\textwidth,
trim=50 30 50 95]{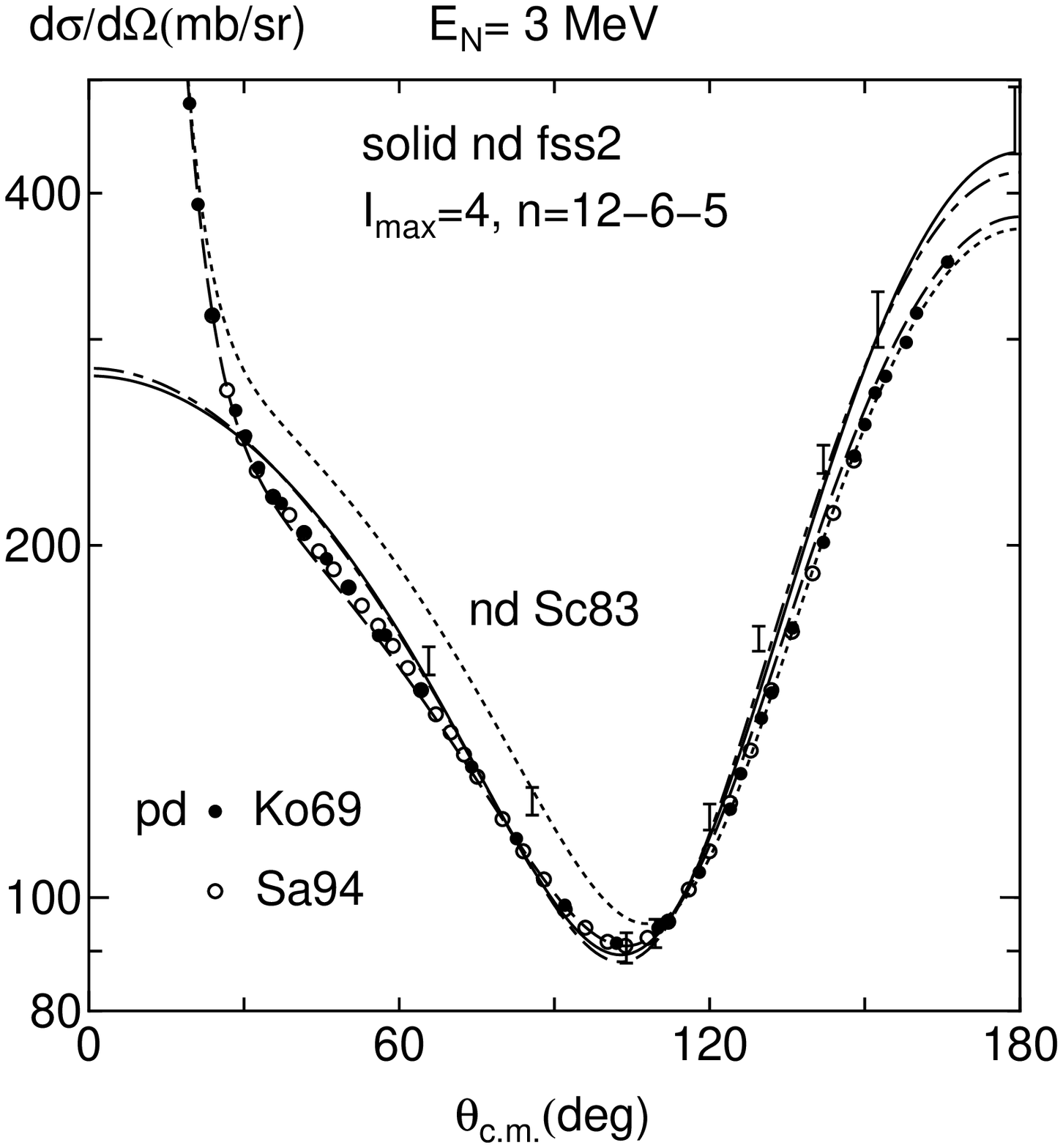}
\end{center}
\label{FDCS-3}
\vspace{-15mm}
\caption{The same as Fig.\,2., but for $E_{N}$=3 MeV. The experimental
data are taken from Ref.\,\citen{Sc83} for Sc83 ($nd$ with errorbars) 
and Ref.\,\citen{Sa94} for Sa94 ($pd$ with empty circles).}
\end{minipage}
\end{figure}

\newpage
\noindent
The Coulomb correction can
be evaluated since the $nd$ and $pd$ eigenphase shifts for $E_{N}=1$, 2 and 3 MeV
by the AV18 potential (or the AV18+UR($3N$) potentials) are both given in Tables
1 and 2 of Ref.\,\citen{Kiev96}, respectively.
The $J$-averaged central phase shifts
calculated from these values imply that the Coulomb modification is significant
(more than $1^{\circ}$) only for the $^{2}S$, $^{4}S$ and $^{4}P$
channels. The difference $ \Delta _{\ell}=[\delta^{C}_{\ell S_{c}}(pd)
-\delta^{C}_{\ell S_{c}}(nd)]_{\rm AV18}$ listed in Table \ref{table3} is
obtained in this way. 
The $pd$ differential cross sections calculated with 
these Coulomb modified nuclear scattering amplitude $f^{N}_{\ell}$ are
plotted in Figs.\,\ref{FDCS-1} -- 4 with dashed curves.
We find excellent agreement with the experimental data for the $pd$
differential cross sections with the aid of the AV18 Coulomb effect.
The overestimation in the case of the original
Coulomb externally corrected approximation is improved mainly by
$7^{\circ}$ -- $10^{\circ}$ modification of the $^{4}S$ phase shift
to the attractive direction (see Table III).
These analyses imply that the correct treatment of the Coulomb force
could reproduce the $pd$ experimental data for the differential cross
sections without reinforcing fss2 with the $3N$ force.

\subsection{The $nd$ effective range parameters}

Since the good correspondence between fss2 and AV18+UR($3N$) are found
in the eigenphase shifts, we can expect that the $S$-wave effective range
parameters for the spin-doublet and quartet channels should also be
reproduced by fss2 without introducing the $3N$ force. These are
determined from Eq.\,(\ref{quartet}) for the quartet channel and
Eq.\,(\ref{doublet}) for the doublet channel.
The $S$-wave effective range parameters are obtained by using Schlessinger's
point method \cite{Sch68} (a type of Pade approximation) to the effective-range
function $K(q_{0})=q_{0}\,{\rm cot}\delta $. This method is convenient to
approximate a function with a pole such as the doublet effective-range
function and to take into account the contributions of the higher order terms in
Eqs.\,(\ref{quartet}) and (\ref{doublet}) in a natural way.
It is very hard in the momentum representation to maintain
sufficient accuracy of the eigenphase shift for the extremely low energies
if we use the realistic deuteron wave function including the $D$-wave component.
We therefore use the sample points of energies between $E_{\rm cm}=200$ keV and
2 MeV shown in Figs.\,5 (the doublet channel) and 6 (the quartet channel)
unless the denominator in the Schlessinger's point method hits zeros.
In order to obtain the eigenphase shifts with 
accuracy of less than $0.01^{\circ}$, we need to take fine mesh points of $p$
for the $NN$ relative motion. We use a set of mesh points $n$=6-10-5 and the
$NN$ partial waves up to $G$-wave ($I_{\rm max}=4$). 
A typical example of fitting to $K(q_{0})$
is shown in Figs.\,5 and 6.
We reconfirm a pole structure existing in the doublet channel at $E_{Q}
=-(3 \hbar^{2}q_{Q}^{2}/4M) \simeq -150$ keV.

\begin{figure}[t]
\begin{minipage}{0.465 \textwidth}
\begin{center}
\includegraphics[scale=0.40, trim=10 150 10 80]{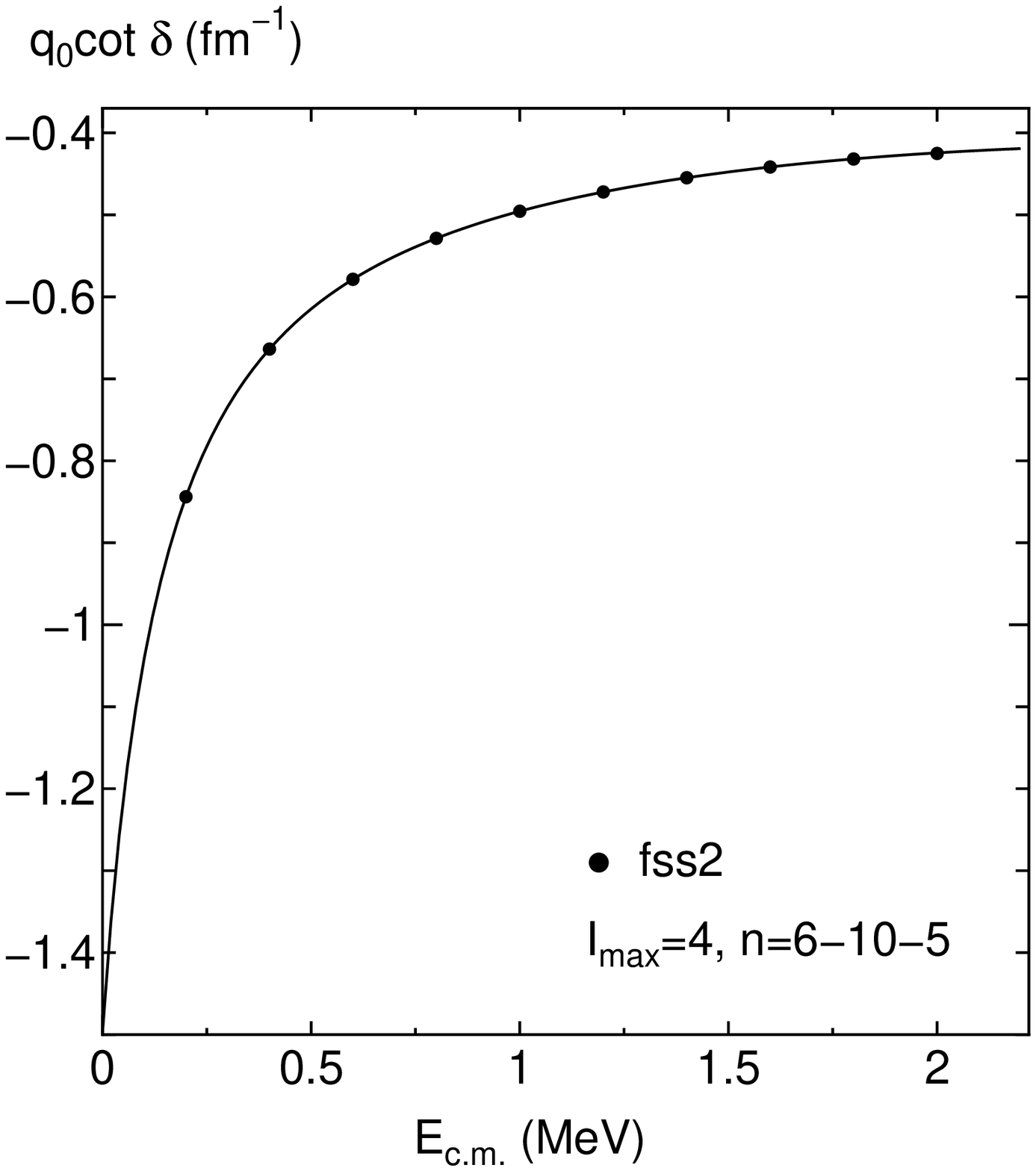}
\caption{The effective-range function $K(q_{0})=q_{0}\cot{\delta}$ for 
the doublet $S$-state as a function of $E_{\rm cm}$.
The curve shows the rational approximation made by the
Schlessinger's point method.\cite{Sch68}}
\end{center}
\end{minipage}
\hspace{5mm}
\begin{minipage}{0.465 \textwidth}
\begin{center}
\includegraphics[scale=0.40, trim=10 150 10 170]{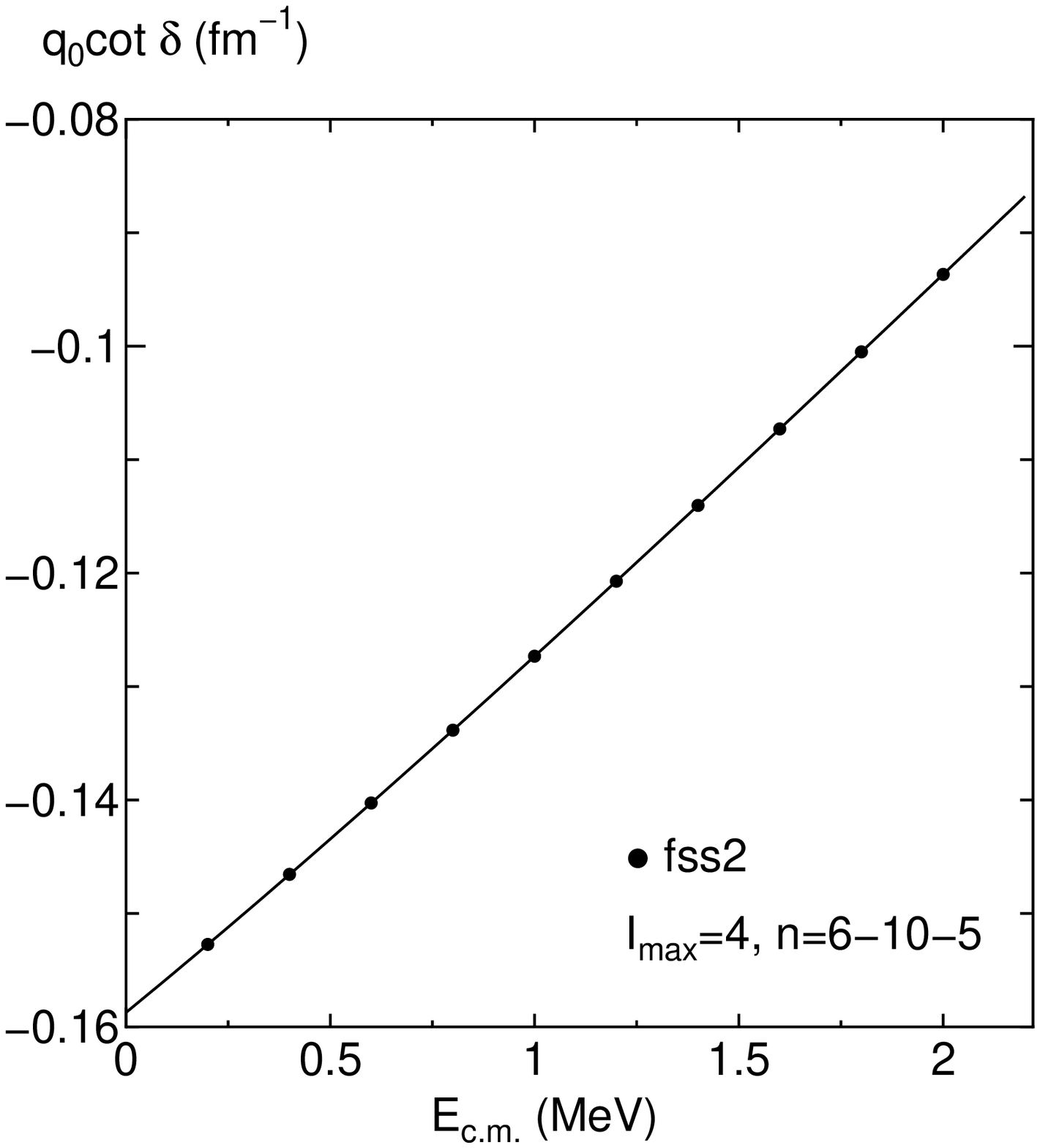}
\caption{The same as Fig.\,5, but $K(q_{0})=q_{0}\cot{\delta}$ for the
quartet $S$-state.}
\end{center}
\end{minipage}
\end{figure}

Table\,\ref{eff1} lists the model-space dependence of the triton binding
energy and effective range parameters. As to the almost converged triton
binding energy $E_{B}({}^{3}\hbox{H})=8.311$ MeV with $I_{\rm max}=6$, 
the small difference from the original result of fss2 (8.326 MeV with
$I_{\rm max}=6$)\cite{triton2} comes mainly from the nonlocal Gaussian
approximation to the interaction kernel, adopted in this study.
Since we have three $NN$ pairs in the triton, this 15 keV difference in the
triton binding energy is consistent with the 4 keV difference in the deuteron
binding energy, which is the difference of 2.2250 MeV (the original result of fss2)
\cite{PPNP} and 2.2206 MeV (the result of the nonlocal Gaussian potential based
on fss2). We find that the spin-quartet scattering length $^{4}a_{nd}$ is quite
insensitive to the model space adopted. This insensitivity is related to
the small distortion effect of the deuteron due to the Pauli repulsion
of the $nd$ interaction in this channel, which is a kinematical constraint
from the spin-isospin quantum numbers. The system is therefore independent
of the details of the $NN$ interaction.
On the other hand, the spin-doublet scattering length $^{2}a_{nd}$
is subject to a strong channel coupling effect as seen in Table \ref{table4}.
We find that the simplest five-channel ($S+D$) calculation, incorporating
the $^{3}S_{1}+{}^{3}D_{1}$ and $^{1}S_{0}$ $NN$ channels only,
yields a value very close to the converged one. However, this is quite
accidental and the well converged value is achieved after many partial
waves, up to the $G$-wave of $NN$ interaction at least, are included.
The values of $|E_{Q}|/(^{2}a_{nd})\sim 220$ keV/fm are almost independent
of the model space. This linear correlation is already suggested in
Ref.\,\citen{Whi76} for the separable potentials.
A strong correlation between $E_{B}({}^{3}\hbox{H})$ and $^{2}a_{nd}$
is also apparent in Table \ref{table4}.

\begin{table}[t]
\caption{The triton binding energy $E_{B}(^{3}\hbox{H})$
and $S$-wave effective range parameters, $^{2}a_{nd}$, $(^{2}r_{e})_{nd}$,
$^{4}a_{nd}$ and $(^{4}r_{e})_{nd}$, predicted by fss2 
for various model spaces with the maximum angular momentum 
of the $NN$ interaction ($I_{\rm max}$) included.
The nonlocal Gaussian potential with 15-point quadrature is used for fss2.
In ``$S+D$'', the $^{3}S_{1}+{}^{3}D_{1}$ and $^{1}S_{0}$ $NN$
channels are only included. The pole energy for the doublet
channel, $E_{Q}$, is also shown. The calculated deuteron binding energy
is 2.2206 MeV. The momentum mesh points with $n$=6-10-5 are used.}
\label{table4}
\renewcommand{\arraystretch}{1.2}
\setlength{\tabcolsep}{2.3mm}
\begin{center}
\begin{tabular}{ccccccc}
\hline \hline
$I_{\rm max}$  & $E_{B}({}^{3}\hbox{H})$ (MeV)
& $^{2}a_{nd}$ (fm) & $(^{2}r_{e})_{nd}$ (fm) & $E_{Q}$ (keV)
& $^{4}a_{nd}$ (fm) & $(^{4}r_{e})_{nd}$ (fm) \\ \hline
$S+D$ & 8.247 & 0.65 & $-149$ & $-147$ & 6.30 & 1.84 \\
1 & 7.948 & 0.94 & $-102$ & $-207$ & 6.31 & 1.85 \\
2 & 8.213 & 0.72 & $-133$ & $-163$ & 6.30 & 1.84 \\
3 & 8.298 & 0.67 & $-146$ & $-151$ & 6.30 & 1.84 \\
4 & 8.307 & 0.66 & $-148$ & $-148$ & 6.30 & 1.84 \\ \hline 
\end{tabular}
\end{center}
\label{eff1}
\end{table}

\begin{table}[t]
\caption{Comparison of the $nd$ scattering lengths, predicted by
using fss2 ($I_{\rm max}=4$), with other models. For the fss2 results,
the charge dependence of the $NN$ force is neglected.
The heading $NN$ implies the calculation using only the $NN$ force,
and $NN$+TM99 the calculation including the Tucson-Melbourne 99 (TM99)
$2 \pi $-exchange $3N$ force \cite{TM99-1,TM99-2}.
The results by CD-Bonn 2000, AV18 and Nijm I for the $NN$ force
are taken from Ref.\,\citen{Wit03} ($I_{\rm max}=5$).
The experimental values are taken from Ref.\,\citen{Di72}.
The values of $^{4}a_{nd}$ are insensitive to the $3N$ force.}
\label{table5}
\renewcommand{\arraystretch}{1.2}
\setlength{\tabcolsep}{4.1mm}
\begin{center}
\begin{tabular}{cccccc}
\hline \hline
model & \multicolumn{2}{c}{$E_{B}(^{3}\hbox{H})$ (MeV)} 
& \multicolumn{2}{c}{$^{2}a_{nd}$ (fm)} & $^{4}a_{nd}$ (fm) \\ \hline
      & $NN$  & $NN$+TM99  & $NN$ & $NN$+TM99 & $NN$(+TM99) \\ \hline                              
fss2  & 8.307 & ----- & 0.66 & ----- & 6.30 \\
\small{CD-Bonn 2000} & 8.005 & 8.482 & 0.925 & 0.569 & 6.347 \\
AV18    & 7.628 & 8.482 & 1.248 & 0.587 & 6.346 \\
Nijm I  & 7.742 & 8.485 & 1.158 & 0.594 & 6.342 \\ \hline
exp & \multicolumn{2}{c}{8.482} & \multicolumn{2}{c}
{$0.65 \pm 0.04$} & $6.35 \pm 0.02$ \\ \hline
\end{tabular}
\end{center}
\label{eff2}
\end{table}

In Table\,\ref{eff2}, we compare $E_{B}({}^{3}\hbox{H})$ and 
$^{2}a_{nd}$ by fss2 with other calculations using
meson-exchange potentials including the $3N$ force.\cite{Wit03}
Here, fss2 does not incorporate the
charge dependence of the $NN$ interaction, but the other calculations
in Table\,\ref{eff2} include this effect. As in $E_{B}(^{3}\hbox{H})$,
it should give an appreciable influence to $^{2}a_{nd}.$
We will estimate the maximum shift of $^{2}a_{nd}$ by simply assuming
the same correlation as Phillips line for the triton binding energy.
For fss2, the slope of the Phillips line, $-0.686$ fm/MeV, yields
0.13 fm for the charge dependence effect of the triton binding energy
190 keV.\cite{Mac89} After the charge dependence correction, $^{2}a_{nd}$ for fss2
would turn out to be $^{2}a_{nd} \sim $ 0.76 -- 0.80 fm.
Table\,\ref{eff2} shows that the effect of the $3N$ force is more
important than the charge dependence of the $NN$ force.
When $NN$ meson-exchange potentials are only used, $^{2}a_{nd}$
is more than 0.9 fm. The experimental values for $^{2}a_{nd}$ and
$E_{B}(^{3}\hbox{H})$ are reproduced only when the $3N$ force is included.
The model fss2 almost reproduces $E_{B}({}^{3}\hbox{H})$ and $^{2}a_{nd}$
simultaneously without introducing the $3N$ force. We should keep in mind
that the mechanism to reproduce $^{2}a_{nd}$ is quite different
from that in the spin-quartet case. The large positive value for
$^{4}a_{nd}$ is related to the Pauli repulsion for the loosely-bound deuteron
cluster. On the other hand, the value of $^{2}a_{nd}$ is entirely from the
dynamical origin related to the fairly large triton binding energy
$E^{\rm exp}_{B}({}^{3}\hbox{H})=8.482$ MeV and the existence of the pole
structure in the effective-range function just below the elastic threshold.
In this special situation, it is natural that more attractive $nd$
interaction, afforded by the $3N$ force in the meson-exchange potentials
or by fss2 without the $3N$ force, can reproduce a smaller
value for $^{2}a_{nd}$, which is the content of the Phillips line.

\begin{table}[t]
\caption{The quartet ($S_{c}=3/2$) and doublet ($S_{c}=1/2$) $S$-wave
contributions to the total cross sections $ \sigma _{\rm tot}$, calculated 
from the effective range parameters for fss2 with $I_{\rm max}=4$ in Table \ref{eff1}.
The calculated total cross sections $ \sigma _{\rm tot}$ are taken from Ref. \citen{ndsc1}.}
\label{table6}
\begin{center}
\renewcommand{\arraystretch}{1.15}
\setlength{\tabcolsep}{3.2mm}
\begin{tabular}{ccccc}
\hline \hline
$E_{n}$ & 0 MeV & 1 MeV & 2 MeV & 3 MeV \\ \hline
$^{4}S$ (mb) & 3325 (99.5\%) & 2055 (72\%)& 1467 (59\%)
& 1128 (54\%) \\ 
$^{2}S$ (mb) & 18.2 (0.54\%)  & 122.4 (4.3\%) & 158.4 (6.3\%)
& 165.5 (7.9\%) \\
$S$-wave  (mb) & 3343 & 2177 (77\%) & 1625 (66\%)
& 1294 (62\%) \\
$ \sigma _{\rm tot}$ (mb) & 3185 (0.15 MeV) & 2833 & 2480 & 2104 \\ \hline
exp. (mb) & $3120 \pm 180$ \cite{Phi80} & $2893.6 \pm 18.2$ \cite{Cl72}
& $2550.6 \pm 11.1$ \cite{Cl72} & $2158.0 \pm 7.2$ \cite{Cl72} \\ 
&(0.07 MeV) & $2854 \pm 39$ \cite{Seg55} & $2537 \pm 10$ \cite{Phi80}
& $2240 \pm 90$ \cite{Nuc46} \\
& & $3110 \pm 200$ \cite{Nuc46} & $2600 \pm 80$ \cite{Nuc46}
& $2160 \pm 86$ \cite{Wil64} \\ \hline
\end{tabular}
\end{center}
\label{eff3}
\end{table}

In Table \ref{eff3}, we show the $S$-wave contributions from the quartet
($S_{c}=3/2$) and doublet ($S_{c}=1/2$) channels to the total cross
sections, calculated from the effective range parameters for fss2 with
$I_{\rm max}=4$ in Table \ref{eff1}. We find that the total cross
sections are dominated by the $S$-wave contribution, which is more than
$60\%$ even for $E_{n}=3$ MeV. Furthermore, the quartet
state is far more important than the doublet state due to the
small values of $|q_{0}\cot{\delta}|$, even considering the statistical
weight factor $(2S_{c}+1)$. As the energy increases, the contribution from
the doublet state becomes appreciable owing to avoiding the pole structure
just below the elastic threshold, but is still less than $10\%$ at
$E_{n}$=3 MeV. This implies a very special situation that an extra
attraction to the $^{2}S_{1/2}$ state by the $3N$ force is unimportant to
reproduce the differential cross sections of the low-energy $nd$
scattering, and they are mainly determined by the magnitude of the
repulsive $^{4}S_{3/2}$ eigenphase shift.

\section{Summary}
Motivated by the success of the QM baryon-baryon interaction
fss2 reproducing the triton binding energy almost correctly without the
$3N$ force,\cite{triton2} we have extended the Faddeev calculation to the low-energy
$nd$ scattering by employing a new algorithm \cite{ndsc1}
to solve the Alt-Grassberger-Sandahs equation \cite{AGS}. The QM $NN$
interaction, formulated in the two-cluster RGM,
has rich contents of nonlocality that the standard meson-exchange
potentials do not possess.\cite{PPNP} In addition to the dominant
nonlocality from the RGM interaction kernel, an extra nonlocality emerges
from the off-shell transformation to eliminate the energy dependence
connected to the normalization kernel,\cite{Su08} which is sometimes neglected in
similar works.\cite{Ju02,Ga07} To reduce the computation time for
three-body calculations, we have developed and used the nonlocal Gaussian
potential constructed from the model fss2.\cite{Fuk09} The nonlocality
and the energy dependence of fss2 is, however, strictly
preserved in this potential model, resulting in an almost the same amount
of the triton binding energy with only 15 keV less.

In this paper, we have examined the effective range parameters of the
$nd$ scattering and the low-energy differential cross sections below
the deuteron breakup threshold. The elastic scattering amplitudes in the
channel-spin representation are parameterized by the eigenphase shifts 
and mixing parameters,\cite{Sey69} from which the $S$-wave effective
range parameters are derived by employing the single-channel effective
range formula. We have reconfirmed that the improved single-channel
effective range formula should be used for the channel-spin doublet
($S_{c}=1/2$) state, to incorporate the pole structure of the
effective-range function, existing just below the elastic threshold.
\cite{Del60,Rei69,Phi69,Whi76} The predicted effective range parameters
by fss2 are: $^{2}a_{nd}=0.66$~fm, $(^{2}r_{e})_{nd}\sim-150$~fm,
$E_{Q}=-(3 \hbar^{2}q_{Q}^{2}/4M)\sim-150$~keV, and $^{4}a_{nd}=6.30$~fm,
$(^{4}r_{e})_{nd}=1.84$~fm without the charge dependece of the $NN$
interaction.
After the charge dependence correction, $^{2}a_{nd}$ for fss2
would turn out to be $^{2}a_{nd} \sim $ 0.76 -- 0.80 fm.
It is found that the almost same $^{2}a_{nd}$
value is accidentally obtained in the restricted model space involving only
the $^{3}S_{1}+{}^{3}D_{1}$ and $^{1}S_{0}$ $NN$ interaction.
However, the deuteron distortion effect in the spin doublet channel 
is so strong that the sufficient partial waves, up to the $G$-wave at least, 
are necessary to obtain the converged result. On the other hand, the
positive value ${}^{4}a_{nd} \sim 6.3$ fm, implying the repulsive nature
of the $nd$ interaction in the spin quartet ($S_{c}=3/2$) channel, is
quite insensitive to the expansion of the model space, owing to the
kinematical constraint by the effect of the Pauli principle.  

A detailed comparison of the $nd$ eigenphase shifts predicted by fss2 and
by the AV18 plus the UR $3N$
potential \cite{Kiev96} shows a prominent resemblance especially for the
$^{2}S_{1/2}$ state. The UR $3N$ potential gives a sizable effect 
only on the $J^{\pi}=1/2^{+}$ channel, while very small effects on the
$^{4}S_{3/2}$ state and the other partial waves. 
We find reasonable agreement with the $nd$ experimental data
for the low-energy differential cross sections.
Since the $pd$ data are more precise
than the $nd$ data, we have also examined the differential cross
sections of the $pd$ elastic scattering. 
We have employed a simple prescription adding
the Coulomb amplitude to the $nd$ scattering amplitudes with the factor 
$e^{i(\sigma _{\ell}+\sigma _{\ell'})}$, 
which is called the Coulomb externally corrected approximation.
\cite{Dol82,De05} The assumption using the $nd$
scattering amplitude with no modification gives too large differential
cross sections for $E_{p}=1$ -- 3 MeV, which implies that the Coulomb
modification of the nuclear phase shifts is very important in this
low-energy region. The Coulomb modified nuclear phase shifts are evaluated
by adding the major difference of the $nd$ and $pd$ eigenphase
shifts predicted by the AV18 potential.\cite{Kiev96}. The modification is
carried out with respect to the $J$-averaged central phase shifts only for
the $^{2}S$, $ \hbox{}^{4}S$ and $^{4}P$ channels. We have found
agreement with the experimental data for the $pd$ differential cross sections
at $E_{p}=1$, 2 and 3 MeV. Since the $^{4}S_{3/2}$ contribution is dominant
in the differential cross sections below the deuteron breakup threshold,
the $^{4}S_{3/2}$ eigenphase shift of the $nd$ (and probably $pd$) elastic
scattering is properly predicted by fss2.

Based on these analyses, we can conclude that the spin-doublet low-energy
eigenphase shift predicted by fss2 is sufficiently attractive to reproduce
predictions of the AV18 plus Urbana $3N$ force, yielding the observed
value of the doublet scattering length and the correct $nd$ and $pd$
differential cross sections below the deuteron breakup threshold. These results
are in accordance with the bound-state calculation of the triton, in which fss2
predicts a nearly correct binding energy close to the experimental value
without introducing the $3N$ force.\cite{triton2} It is important to examine
if the correct treatment of the Coulomb force for fss2 can reproduce the $pd$
differential cross sections and the polarization observables below the
deuteron breakup threshold.

\section*{Acknowledgments}
The authors would like to thank Professor B. F. Gibson and Professor
S. Ishikawa for informing them on the pole structure of the spin-doublet
effective-range function. They also thank Professor K. Sagara for
providing them the $pd$ experimental data taken by the Kyushu University
group. This work was supported by the Grant-in Aid for Scientific
Research on Priority Areas (Grant No. 2002803), and by the Grant-in-Aid
for the Global COE Program ``The Next Generation of Physics, Spun from
Universality and Emergence" from the Ministry of Education, Culture,
Sports, Science and Technology (MEXT) of Japan. It was also supported by
the core-stage backup subsidies of Kyoto University. The numerical
calculations were carried out on Altix3700 BX2 at YITP in Kyoto University.

\appendix
\section{The spin-isospin factors and the rearrangement coefficients in
Eq.~(\ref{fad6})}
In this appendix, 
we give the rearrangement coefficients $g_{\gamma,\gamma'}(q,q',x)$
in Eq. (\ref{fad6}) for the permutation operator $P=P_{(123)}+P_{(123)}^{2}$,
together with the spin-isospin factors for the $3N$ system.
The two contributions from $P_{(123)}=P_{(12)}P_{(23)}$ and
$P_{(123)}^{2}=P_{(13)}P_{(23)}$ become equal owing to the Pauli principle
for the particle 1 and particle 2. We therefore only need to calculate the
rearrangement coefficients for $(-2)P_{(13)}$. These are given by
\vspace{-1.5mm}
\begin{equation}
g_{\gamma,\gamma'}(q,q',x)=
\sum _{\lambda _{1}+\lambda _{2}=\lambda}\sum
_{\lambda'_{1}+\lambda'_{2}=\lambda'}q^{\lambda'_{1}+\lambda _{2}}
q'^{\lambda _{1}+\lambda'_{2}}
\left(\dfrac{1}{2}\right)^{\lambda _{2}+\lambda'_{2}}
\sum _{k}(2k+1)g^{\lambda _{1}\lambda'_{1}k}_{\gamma,\gamma'}P_{k}(x)~,
\end{equation}
with
\begin{equation}
g^{\lambda _{1}\lambda'_{1}k}_{\gamma,\gamma'}=
\sum _{LS}(X_{N})^{LSJ}_{\gamma,\gamma'}
G^{\lambda _{1}\lambda'_{1}kL}_{(\lambda \ell),(\lambda' \ell')}~,
\label{gXG}
\end{equation}
and $P_{k}(x)$ being Legendre polynomials.
The spatial rearrangement factor $G^{\lambda _{1}\lambda'_{1}kL}_
{(\lambda \ell),(\lambda' \ell')}$ is given by
\begin{eqnarray}
&& G^{\lambda _{1}\lambda'_{1}kL}_{(\lambda \ell),(\lambda' \ell')}
=G^{\lambda'_{1}\lambda _{1}kL}_{(\lambda' \ell'),(\lambda \ell)}
=4 \pi \left[\frac{(2 \lambda+1)!(2 \lambda'+1)!}{(2 \lambda _{1}+1)!
(2 \lambda _{2}+1)!(2 \lambda'_{1}+1)!(2 \lambda'_{2}+1)!}\right]^{\frac{1}{2}}
\nonumber \\
&& \quad \times \int d \widehat{\bm q} d \widehat{\bm q}'
\left[Y_{(\lambda _{1}\lambda _{2})\lambda}(\widehat{\bm q}',\widehat{\bm q})
Y_{\ell}(\widehat{\bm q})\right]^*_{LM}
P_{k}(\widehat{\bm q}\cdot \widehat{\bm q}')
\left[Y_{(\lambda'_{1}\lambda'_{2})\lambda'}(\widehat{\bm q},\widehat{\bm q}')
Y_{\ell'}(\widehat{\bm q}') \right]_{LM} \nonumber \\ 
&& =\left[\frac{(2 \lambda+1)!(2 \lambda'+1)!}{(2 \lambda _{1})!
(2 \lambda _{2})!(2 \lambda'_{1})!(2 \lambda'_{2})!}\right]^{\frac{1}{2}}
\widehat{\lambda}\widehat{\ell}\widehat{\lambda'}\widehat{\ell'}\sum _{ff'}
\langle \lambda _{2}0 \ell 0|f0 \rangle \langle \lambda'_{2}0 \ell' 0|f'0
\rangle \langle k0 \lambda _{1}0|f'0 \rangle \nonumber \\
&& \quad \times \langle k0 \lambda'_{1}0|f0 \rangle
\begin{Bmatrix}
f & L & \lambda _{1} \\
\lambda  & \lambda _{2} & \ell 
\end{Bmatrix}
\begin{Bmatrix}
f' & L & \lambda'_{1} \\
\lambda' & \lambda'_{2} & \ell' 
\end{Bmatrix}
\begin{Bmatrix}
\lambda'_{1} & f' & L \\
\lambda _{1} & f & k
\end{Bmatrix}~.
\end{eqnarray}
Here, $ \widehat{\lambda}=\sqrt{2 \lambda+1}$ etc. and
$ \lambda _{2}=\lambda-\lambda _{1}$, $ \lambda'_{2}=\lambda'-\lambda'_{1}$
with $ \lambda _{1}=0$ -- $\lambda $, $ \lambda'_{1}=0$ -- $ \lambda'$.
The explicit expression of the spin-isospin factors
$(X_{N})^{LSJ}_{\gamma,\gamma'}$ depends on the channel specification scheme.
For the $LS$ coupling scheme
\begin{eqnarray}
\langle{\widehat{\bm p}},{\widehat{\bm q}};123|\gamma \rangle=
\left[Y_{(\lambda \ell)L}({\widehat{\bm p}}, {\widehat{\bm q}})
\left[\chi _{st}(1,2)\chi _{\frac{1}{2}\frac{1}{2}}(3)\right]_{SS_{z};
\frac{1}{2}T_{z}}\right]_{JJ_{z}}\ ,
\end{eqnarray}
with $ \gamma=[(\lambda \ell)L(s{\scriptstyle{\frac{1}{2}}})S]
JJ_{z};(t {\scriptstyle{\frac{1}{2}}}){\scriptstyle{\frac{1}{2}}}T_{z}~$,
we do not need the $LS$ sum in Eq.~(\ref{gXG}), since the orbital angular
momentum $L$ and the total spin $S$ are both conserved. We modify 
Eq.~(\ref{gXG}) to
\begin{eqnarray}
(X_{N})^{LSJ}_{\gamma,\gamma'}\rightarrow
\left(X^{S{\scriptstyle{\frac{1}{2}}}}_{N}\right)^{LSJ}_{st,s't'}
\equiv-2X^{S}_{s,s'}X^{{\scriptstyle{\frac{1}{2}}}}_{t,t'}~,
\end{eqnarray}
where a common definition of the spin and isospin factors
\begin{equation}
X^{\frac{3}{2}}_{s,s'}=
\Biggl(\begin{array}{cc}
0 & 0 \\ 0 & 1 
\end{array}\Biggr)
,\quad X^{\frac{1}{2}}_{s,s'}=
\begin{pmatrix}
\frac{1}{2} & -\frac{\sqrt{3}}{2} \\ -\frac{\sqrt{3}}{2} & -\frac{1}{2}
\end{pmatrix}~,
\label{ApA6}
\end{equation}
is used. In Eq.~(\ref{ApA6}), the upper row (or the left-most column)
corresponds to $s$=0 ($s'$=0) and the second row (or the right-most column)
corresponds to $s$=1 ($s'$=1). For the $jj$-coupling scheme 
\begin{eqnarray}
\langle{\widehat{\bm p}},{\widehat{\bm q}};123|\gamma \rangle
=\left[[Y_{\lambda}({\widehat{\bm p}})\chi _{st}(1,2)]_{I}
[Y_{\ell}({\widehat {\bm q}})\chi _{{\scriptstyle{\frac{1}{2}}},
{\scriptstyle{\frac{1}{2}}}}(3)]_{j}\right]_{JJ_{z};
{\scriptstyle{\frac{1}{2}}}T_{z}}\ ,
\end{eqnarray}
with $ \gamma=[(\lambda s)I(\ell{\scriptstyle{\frac{1}{2}}})j]JJ_{z};
(t{\scriptstyle{\frac{1}{2}}}){\scriptstyle{\frac{1}{2}}}~T_{z}$,
we use the superposition of Eq.~(\ref{gXG}) with
\begin{equation}
(X_{N})^{LSJ}_{\gamma,\gamma'}=
\begin{bmatrix}
\lambda & s & I \\
\ell & {\scriptstyle{\frac{1}{2}}} & j \\
L & S & J
\end{bmatrix}
\begin{bmatrix}
\lambda' & s '& I' \\
\ell' & {\scriptstyle{\frac{1}{2}}} & j' \\
L & S & J
\end{bmatrix}
\left(X^{S{\scriptstyle{\frac{1}{2}}}}_{N}\right)^{LSJ}_{st,s't'}~.
\end{equation}
The coefficients in the channel-spin formalism 
with Eq.~(\ref{LS-couple}) are similarly obtained as
\begin{eqnarray}
(X_{N})^{LSJ}_{\gamma,\gamma'} &=& \sum _{jj'}
\begin{bmatrix}
0 & \ell & \ell \\
I & {\scriptstyle{\frac{1}{2}}} & S_{c} \\
I & j & J \\
\end{bmatrix}
\begin{bmatrix}
\lambda & s & I \\
\ell & {\scriptstyle{\frac{1}{2}}} & j \\
L & S & J
\end{bmatrix}
\begin{bmatrix}
0 & \ell' & \ell' \\
I' & {\scriptstyle{\frac{1}{2}}} & S'_{c} \\
I' & j' & J \\
\end{bmatrix}
\begin{bmatrix}
\lambda' & s '& I' \\
\ell' & {\scriptstyle{\frac{1}{2}}} & j' \\
L & S & J
\end{bmatrix} \nonumber \\
&& \times \left(X^{S{\scriptstyle{\frac{1}{2}}}}_{N}\right)^{LSJ}_{st,s't'}~.
\end{eqnarray}
For the practical calculations, it is convenient first to calculate the
product of one 6-$j$ and one 9-$j$ coefficients defined through
\begin{equation}
A^{LSJ}_{(\ell S_{c})(\lambda s)I}=\sum _{j}
(-1)^{I+\ell+S_{c}+j}\widehat{S_{c}}\widehat{j}
\begin{Bmatrix}
{\scriptstyle{\frac{1}{2}}} & S_{c} & I \\
J & j & \ell
\end{Bmatrix}
\begin{bmatrix}
\lambda & s & I \\
\ell & {\scriptstyle{\frac{1}{2}}} & j \\
L & S & J
\end{bmatrix}~.
\end{equation}
Furthermore, the factor 2 will be taken off to cancel with the 1/2
factor over the $x$ integral. Then we use for Eq.~(\ref{gXG})
\begin{equation}
\frac{1}{2}g^{\lambda _{1}\lambda'_{1}k}_{\gamma,\gamma'}
=-X^{\scriptstyle{\frac{1}{2}}}_{t,t'}\sum _{LS}
A^{LSJ}_{(\ell S_{c})(\lambda s)I}A^{LSJ}_{(\ell' S'_{c})
(\lambda' s')I'}X^{S}_{s,s'}
G^{\lambda _{1}\lambda'_{1}kL}_{(\lambda \ell),(\lambda' \ell')}~.
\end{equation}

\end{document}